\documentclass[prd,twocolumn,floatfix,preprintnumbers,letterpaper]{revtex4}
\usepackage{amssymb}
\usepackage{graphics}
\usepackage{epsfig}
\usepackage{amsmath}

\def\4He{$^4$He}

\begin{document}

\title{Do Fermions and Bosons Produce the Same Gravitational Field?}
\author{John D. Barrow}
\affiliation{DAMTP, Centre for Mathematical Sciences, Cambridge University, Cambridge CB3
0WA}
\author{Robert J. Scherrer}
\affiliation{Department of Physics and Astronomy, Vanderbilt University, Nashville,
TN~37235}

\begin{abstract}
We examine some cosmological consequences of gravity coupling with different
strength to fermions and bosons. We show that this leads to a different
perturbation of the standard picture of primordial nucleosynthesis than the
addition of extra neutrino types or overall scaling of the value of $G$.
Observed abundances of deuterium and $^4$He place bounds on the ratio of the
bosonic gravitational constant ($G_B$) to the fermionic gravitational
constant ($G_{F}$) of $0.45<{G_{B}}/{G_{F}}<0.92$ at $1\sigma $ and $%
0.33<G_B/G_F<1.10~~$at $2\sigma $. A value of $G_B < G_F$ can reconcile the
current ``tension" between the abundances of deuterium and $^4$He predicted
by primordial nucleosynthesis. We comment briefly on other cosmological
effects.
\end{abstract}

\maketitle

We examine some cosmological consequences of assuming that gravity is not
blind to statistics and that the gravitational coupling constant, $G,$ is
different for fermions and bosons. The gravitational constant is notoriously
difficult to measure to high precision and is poorly known in comparison to
other fundamental constants \cite{CODATA, Scherrer, div}. Hence there has
been extensive experimental and theoretical exploration of possible
deviations from the standard Newtonian and general relativistic theory of
the gravitational interaction \cite{fis}. These investigations have recently
been rejuvenated by the realisation that \textquotedblleft large" extra
dimensions can be accommodated in string theories and can lead to anomalous
gravitational interactions on scales as large as $10^{-5}$ m \cite{dim, sd}.
Supersymmetry assumes that there is a fundamental symmetry between fermions
and bosons but this symmetry must have been broken at the TeV scale. Perhaps
this breaking produces further asymmetries in fermion-boson properties or
couplings at lower energies?

In allowing $G$ to be different for bosons and fermions, the key question is the
scale over which a particle is considered a boson or a fermion.  We have chosen
to set this scale equal to the nucleon scale, i.e., all nucleons are considered
``fermions" in our discussion.  There are certainly alternatives to this
approach.  For example, one could treat the constituents of the nucleons as the
fundamental particles, with the quarks coupling to $G$ as fermions and the gluons
as bosons; this is certainly the correct approach prior to the quark-hadron
phase transition.  Alternatively, one could define bosons and fermions on a larger
scale, so that a helium nucleus, for instance, would couple to gravity as a boson
rather than as four fermions.  We believe that our approach to this
admittedly speculative topic is the most logical,
but it is important to note that other definitions of bosons and
fermions would yield other results.

Note that we assume that this differential coupling of bosons and fermions to gravity
affects only the source term in the Einstein equations.  We assume that
the Equivalence Principle still holds, so that a boson and a fermion in
a given gravitational field will still follow exactly the same trajectories.
Because of this distinction,
the gravitational coupling to bosons is not probed experimentally. All
direct terrestrial experimental measurements of $G$ are made with a
fermionic source (see ref. \cite{fis} for a review), so that the measured $G$
is really $G_{F}$, the coupling to fermions, and the bosonic $G$, denoted 
$G_{B}$, is undetermined. Observations of the gravitational deflection of
light do not isolate $G_{B}$ either, since they measure the behavior of
bosons under a fermionic source, and under the assumptions noted above, a
boson in the gravitational field generated by a fermion will follow the same
trajectory as a fermion in this gravitational field.

The Shapiro time-delay effect has been
considered as a constraint on Equivalence Principle violation under the
assumption that gravity couples differently to neutrinos and photons in the
PPN approximation. A comparison of the difference in arrival times for
neutrinos and photons from SN1987A allows a bound to be placed on any PPN $%
\gamma $ parameter difference between photons and neutrinos, with $%
\left\vert \gamma _{photon}-\gamma _{neutrino}\right\vert <O(10^{-3}),$ \cite%
{longo, kr}, with small uncertainties due to the gravitational field
of the Galaxy. \ Differences in gravitational coupling to $\nu _{e}$ and $%
\bar{\nu}_{e}$ have also been studied for SN1987A, by Pakvasa et al \cite%
{pak}, yielding $\left\vert \gamma _{\nu _{e}}-\gamma _{\bar{\nu}%
_{e}}\right\vert <10^{-6}$. It is interesting to note that a difference only
of order $10^{-14}$ in the coupling of gravity to $\nu _{e}$ \ and $\nu
_{\mu }$ would make the gravitational transformation of $\nu _{e}$ to $\nu
_{\mu }$ of similar strength to the conversion due to the MSW effect \cite%
{wolf, mik, fis}. In what follows we will consider the simplest scenario of
a difference in gravitational coupling constant for bosons and fermions, but
assume that the coupling is the same for particles and antiparticles and for
all fermionic species. These assumptions can be relaxed straightforwardly if
required.  (See, e.g., Ref. \cite{MR1} for differential coupling to different
families, and Ref. \cite{MR2} for different coupling to particles and antiparticles).

The only place where bosons act as a significant source term for gravity is
in the overall expansion of the universe as a whole, in which photons
contribute significantly (at least in the early universe) to the overall
energy density. In seeking to constrain $G_B$, therefore, the best place to
look is at early universe tests such as Big Bang Nucleosynthesis (BBN) and
the cosmic microwave background (CMB). We will explore the former in detail
and comment briefly on the latter.

In the standard flat Friedmann cosmological model, the overall Hubble
expansion rate is given at early times by 
\begin{equation}
H=\left( \frac{8}{3}\pi G\rho \right) ^{1/2},  \label{expand}
\end{equation}%
where $\rho $ is the total gravitating density. Now consider a model in
which we have two separate gravitational constants: $G_{F}$ for fermions and 
$G_{B}$ for bosons. In our model, $G_{F}$ is just set equal to the measured
value of Newton's constant, $G=6.6742\pm 0.0010\times 10^{-8}$ cm$^{3}$ gm$%
^{-1}$ s$^{-2}$ \cite{CODATA}, so all of the information in the model is
contained in the ratio $G_{B}/G_{F}$, which we will denote by $f_{BF}$: 
\begin{equation}
f_{BF}\equiv \frac{G_{B}}{G_{F}}.
\end{equation}%
Then equation (\ref{expand}) for the expansion rate becomes 
\begin{eqnarray}
H &=&\left( \frac{8}{3}\pi (G_{B}\rho _{B}+G_{F}\rho _{F})\right) ^{1/2}, \\
&=&\left( \frac{8}{3}\pi G(f_{BF}\rho _{B}+\rho _{F})\right) ^{1/2},
\end{eqnarray}%
and the problem is equivalent to changing the density of bosons by the
factor of $f_{BF}$ and leaving the fermion density unchanged.

This model is qualitatively similar to both adding (or subtracting)
relativistic degrees of freedom, or to changing the overall value of $G$,
both of which have been extensively explored in connection with BBN, but we
will now show that it is different from either of these previously studied
situations.

Consider first the effect on the expansion rate of adding additional
relativistic degrees of freedom to $\rho $. (The resulting variation in the
primordial element production can be used to constrain such a change, e.g.,
Refs. \cite{HT, sh, SSG,Kneller1,Barger,Kneller2}). It is conventional to
parametrize such a change in terms of the effective number of additional
(i.e., beyond $\nu _{e},\nu _{\mu },\nu _{\tau }$) relativistic
two-component neutrino degrees of freedom, $\Delta N_{\nu }$. In this case
the total relativistic energy density prior to $e^{+}e^{-}$ annihilation is 
\begin{equation}
\rho =[2+(7/8)10+(7/8)2\Delta N_{\nu }]\frac{\pi ^{2}}{30}T^{4},
\label{rho1}
\end{equation}%
where the first term in square brackets counts the boson degrees of the
freedom (photons), the second counts the fermionic degrees of freedom ($%
e^{+}e^{-}$ and $\nu \bar{\nu}$), the third counts any hypothetical
additional relativistic degrees of freedom, and $T$ is the photon
temperature. (We use units with $\hbar
=c=k_{B}=1$ throughout). After $e^{+}e^{-}$ annihilation, when the photons
are heated relative to the neutrinos, the corresponding energy density is 
\begin{equation}
\rho =[2+(7/8)(4/11)^{4/3}6+(7/8)(4/11)^{4/3}2\Delta N_{\nu }]\frac{\pi ^{2}%
}{30}T^{4}.  \label{rho2}
\end{equation}%
These expressions can be rewritten in terms of the photon density, $\rho
_{\gamma }$, to give \cite{Kneller2} 
\begin{eqnarray}
\rho _{before} &=&5.375[1+0.1628\Delta N_{\nu }]\rho _{\gamma },
\label{delta1} \\
\rho _{after} &=&1.681[1+0.1351\Delta N_{\nu }]\rho _{\gamma },
\label{delta2}
\end{eqnarray}%
where the subscripts $before$ and $after$ refer to the density before and
after $e^{+}e^{-}$ annihilation, respectively.

Another way to parametrize the change in the expansion rate is through a
change in the overall value of $G$, or, equivalently, multiplying equation (%
\ref{expand}) by a \textquotedblleft speed-up" factor $S$. By incorporating
such a change into BBN, the resulting changes in the element abundances can
be used to constrain a time-shift in the value of $G$ \cite%
{Barrow,Yang,Accetta,Copi}. As emphasized by Kneller and Steigman \cite%
{Kneller2}, these two ways of modifying BBN (adding additional relativistic
degrees of freedom, or multiplying $G$ by a constant) while qualitatively
similar, are inequivalent. This can be most easily seen by noting that
changing $G$ by some fixed factor $f_{G}$ is completely equivalent to
changing $\rho $ by this same factor; the \textquotedblleft effective" $\rho 
$ which enters into equation $(\ref{expand})$ is then just 
\begin{eqnarray}
\rho _{before} &=&5.375[f_{G}]\rho _{\gamma },  \label{G1} \\
\rho _{after} &=&1.681[f_{G}]\rho _{\gamma }.  \label{G2}
\end{eqnarray}%
In order to make a change in $G$ equivalent to a change in the number of
effective neutrino degrees of freedom, we would need $f_{G}=1+0.1628\Delta
N_{\nu }$ before $e^{+}e^{-}$ annihilation and $f_{G}=1+0.1351\Delta N_{\nu }
$ after $e^{+}e^{-}$ annihilation; obviously, it is impossible to satisfy
both equations simultaneously.

Now consider the equivalent expressions in our model, when $G_{B}\neq G_{F}$%
. In equations (\ref{rho1}-\ref{rho2}), the first term in brackets gives the
contribution to the energy density from the photons, which are the only
bosonic degrees of freedom present during BBN. Hence, when $G_{B}\neq G_{F}$%
, the effective density becomes 
\begin{eqnarray}
\rho _{before} &=&[2f_{BF}+(7/8)10]\frac{\pi ^{2}}{30}T^{4}, \\
\rho _{after} &=&[2f_{BF}+(7/8)(4/11)^{4/3}6]\frac{\pi ^{2}}{30}T^{4}.
\end{eqnarray}%
These expressions can be rewritten in terms of the photon density as 
\begin{eqnarray}
\rho _{before} &=&[f_{BF}+4.375]\rho _{\gamma },  \label{f1} \\
\rho _{after} &=&[f_{BF}+0.681]\rho _{\gamma }.  \label{f2}
\end{eqnarray}%
A comparison of equations (\ref{f1}) - (\ref{f2}) with equations (\ref%
{delta1}) - (\ref{delta2}) and with equations (\ref{G1}) - (\ref{G2}) shows
that changing $G_{B}/G_{F}$ is inequivalent (in terms of its effect on the
expansion rate) to changing either the overall value of $G$, or adding
additional relativistic degrees of freedom. This can be seen most easily by
fixing $\rho _{before}$ to the same multiple of $\rho _{\gamma }$ in all
three cases; it is easy to see that the resulting values of $\rho _{after}$
are all different.

We now consider the effect of taking $f_{BF}\neq 1$ on the primordial
element abundances. The primordial production of \4He is controlled by the
competition between the expansion rate and the rates for the weak
interactions which govern the interconversion of neutrons and protons: 
\begin{eqnarray}
n+\nu _{e} &\leftrightarrow &p+e^{-},  \notag \\
n+e^{+} &\leftrightarrow &p+\bar{\nu}_{e},  \notag \\
n &\leftrightarrow &p+e^{-}+\bar{\nu}_{e}.
\end{eqnarray}%
At high temperatures, $T\gtrsim 1$ MeV, the weak-interaction rates are
faster than the expansion rate, $H$, and the neutron-to-proton ratio ($n/p$)
tracks its equilibrium value $\exp [-\Delta m/T]$, where $\Delta m$ is the
neutron-proton mass difference. As the universe expands and cools, the
expansion rate becomes too fast for the kinetic equilibrium to be maintained
by weak interactions and $n/p$ freezes out. Nearly all the neutrons which
survive this freeze-out are converted into \4He as soon as deuterium becomes
stable against photodisintegration, but trace amounts of other elements are
produced, including deuterium (see, e.g., Ref. \cite{Olive} for a review).
Therefore, the primordial production of \4He is very sensitive to the
expansion rate of the Universe at temperatures $\sim $1 MeV, so BBN has been
used many times to constrain any change in this expansion rate.

As is the case for other models which change the expansion rate, the
primordial deuterium abundance is most sensitive to changing the
baryon-photon ratio $\eta $, and it essentially provides the upper and lower
bounds on $\eta $. The predicted abundance of \4He within this range for $%
\eta $ can then be calculated as $f_{BF}$ varies, allowing bounds to be
placed on $f_{BF}$. (This is something of an oversimplification, as the
deuterium abundance also depends weakly on $f_{BF}$; our calculation
correctly incorporates this dependence).

The primordial abundance of deuterium has been inferred from QSO absorption
systems. We use the abundance estimated by Kirkman et al. \cite{Kirkman}: 
\begin{equation}
\log (D/H)=-4.556\pm 0.064,
\end{equation}%
where all errors are quoted at the $1-\sigma $ level.

The abundance of $^{4}$He can be inferred from low-metallicity HII regions,
but there are significant discrepancies between different estimates (see Ref.%
\cite{Barger} for a recent analysis, and references therein). We will follow
Ref. \cite{Barger} and take the primordial $^{4}$He mass fraction, $Y_{P}$,
to be 
\begin{equation}
Y_{P}=0.238\pm 0.005,
\end{equation}%
an estimate which is consistent, for example, with that found in the review
of Olive et al. \cite{Olive}.

Using a modified version of the Kawano nucleosynthesis code \cite{Kawano} we
scan over the ($\eta$, $f_{BF}$) plane, calculating the likelihood for a
given pair of values. (In doing so, we take the distribution of $\log (D/H) $
to be Gaussian as in, e.g., Refs. \cite{Copi,Kirkman}, rather than taking
the distribution of $D/H$ to be Gaussian as in, e.g., Ref. \cite{Barger}. In
practice, this should have only a small effect).  The $1-\sigma$ and $2-\sigma$ contours are
shown in Fig. 1 (where we take $\eta = \eta_{10} \times 10^{-10}$).
\begin{figure}[htb]
\centerline{\epsfxsize=3truein\epsffile{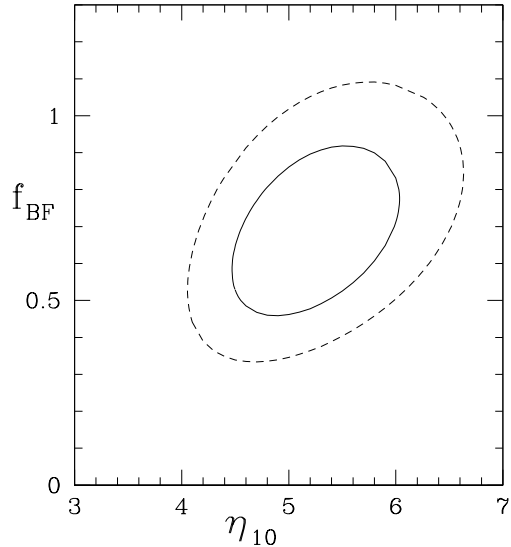}}
\caption{The $1-\sigma$ (solid) and $2-\sigma$ (dashed) contours
in the $\eta_{10}$, $f_{BF}$ plane, where $\eta =  \eta_{10} \times 10^{-10}$
and $f_{BF} = G_B/G_F$.}
\end{figure}
The limits on $f_{BF}$, at the $1-\sigma $ and $2-\sigma $ levels, are: 
\begin{eqnarray}
0.45 &<&f_{BF}<0.92,~~~(1-\sigma ), \\
0.33 &<&f_{BF}<1.10,~~~(2-\sigma ).
\end{eqnarray}%
These limits are the main result of our paper.

Note that the standard model is excluded at the $1-\sigma $ level. This is
not surprising; it is due to the tension which currently exists between the
observed deuterium abundance and the observed $^{4}$He abundance; the former
prefers a higher value of $\eta $ than does the latter. This sort of result,
then, also shows up in discussions of changing $\Delta N_{\nu }$ \cite%
{Barger} and in analyses of other allowed deviations from standard physics
(e.g., Ref. \cite{webb}). Note that a value of $G_{B}$ less than $G_{F}$
provides yet another mechanism for resolving this tension. Clearly, $%
G_{B}\gg G_{F}$ is excluded by BBN, but $G_{B}$ could be much smaller than $%
G_{F}$.

Of course, these conclusions are strongly dependent on the parameters that go into
the calculation.  A higher primordial $^4$He abundance, such as that claimed, for instance,
in Ref. \cite{IT}, would eliminate the tension between the deuterium and $^4$He abundances,
and shift our allowed region upward in Fig. 1.  A similar effect would occur if the neutron
lifetime were measured to be shorter than the currently accepted value \cite{neutron}.

Including CMB data in our analysis will improve these limits, but not by
much. Although, as we have emphasized, our model differs both from changing $%
\Delta N_{\nu }$, or from changing $G$, it is qualitatively similar to both
of these. For the case of changing $\Delta N_{\nu }$, the addition of CMB
data constrains $\eta $ more tightly than BBN alone, but it has only a small
effect on the overall limits on $\Delta N_{\nu }$, compared to using BBN
alone \cite{Barger}. Similarly, BBN provides a stronger constraint on a
change in $G$ than does the CMB (compare, for example, the results in Ref. 
\cite{Scherrer} for BBN with those in Ref. \cite{Zahn} for the CMB).

It is difficult to imagine other environments in which the value of $G_{B}$
could manifest itself. If a scenario of baryon number generation by out of
equilibrium decay of superheavy vector or scalar bosons \cite{wein,FOT} were
precisely established as the source of the value of $\eta$, then there would
be a dependence of $\eta $ on $f_{BF}$ via the ratio of boson decay rate to
the total expansion rate, but this is not the case at present. If the dark
matter were bosonic, then the overall expansion rate would also be altered
in the matter-dominated era. This could only be detected, however, if the
number density and mass of the dark-matter particle were independently
determined, rather than being inferred from the Hubble expansion itself. The
effect on large-scale structure in this case is equivalent to a model with
one gravitational interaction for dark matter and a different one for
luminous matter \cite{gib}. It remains to be seen whether there are further
consequences of $G_{B}\neq G_{F}$ at very early times (when $T>>1$ MeV)
which could lead to observable consequences at low energy.

\acknowledgments

We thank G. Steigman for helpful discussions.  We thank
J. Beacom, T. Jacobson, C. Stubbs, L. Krauss, and E. Masso for helpful comments
on the manuscript.
R.J.S. was supported in part
by the Department of Energy (DE-FG05-85ER40226).


\begin{thebibliography}{99}
\bibitem{CODATA} http://www.nist.gov/

\bibitem{Scherrer} R.J. Scherrer, \prd {\bf 69}, 107302 (2004).

\bibitem{div} G. Dvali, hep-th/0402130.

\bibitem{fis} E. Fischbach and C.L. Talmadge, \textit{The Search for
Non-Newtonian Gravity}, Springer-Verlag, New York, (1999).

\bibitem{dim} G. Dvali, G. Gabadaze and M. Porrati, Mod. Phys. Lett. A 
\textbf{15}, 1717 (2000).

\bibitem{sd} S. Dimopoulos and G. Landsberg, Phys. Rev. Lett. \textbf{87},
161602 (2001).

\bibitem{longo} M.J. Longo, Phys. Rev. Lett. \textbf{60}, 173 (1988).

\bibitem{kr} L. Krauss and S. Tremaine, Phys. Rev. Lett. \textbf{60}, 176
(1988).

\bibitem{pak} S. Pakvasa, W.A. Simmons and T.J. Weiler, Phys. Rev. D \textbf{%
39}, 1761 (1989).

\bibitem{wolf} L. Wolfenstein, Phys. Rev. D \textbf{17}, 2369 (1978).

\bibitem{mik} S.P. Mikheyev and A.Y. Smirnov, Sov. J. Nucl. Phys.\textbf{\ 42%
}, 913 (1985).

\bibitem{MR1} E. Masso and F. Rota, \prd \textbf{68}, 123504 (2003).

\bibitem{MR2} E. Masso and F. Rota, astro-ph/0406660.

\bibitem{HT} F. Hoyle and R.J. Tayler, Nature \textbf{203}, 1108 (1964).

\bibitem{sh} V.F. Shvartsman, Sov. Phys. JETP Lett.\textbf{\ 9}, 184 (1969)

\bibitem{SSG} G. Steigman, D.N. Schramm, and J.E. Gunn, Phys. Lett. B 
\textbf{66}, 202 (1977).

\bibitem{Kneller1} J.P. Kneller, R.J. Scherrer, G. Steigman, and T.P.
Walker, \prd {\bf 64}, 123506 (2001).

\bibitem{Kneller2} J.P. Kneller and G. Steigman \prd {\bf 67}, 063501 (2003).

\bibitem{Barger} V. Barger, J.P. Kneller, H.-S. Lee, D. Marfatia, and G.
Steigman, Phys. Lett. B \textbf{566}, 8 (2003).

\bibitem{Barrow} J.D. Barrow, Mon. Not. R. Astr. Soc. \textbf{184}, 677
(1978).

\bibitem{Yang} J. Yang, D.N. Schramm, G. Steigman, and R.T. Rood, 
\apj {\bf
227}, 697 (1979).

\bibitem{Accetta} F.S. Accetta, L.M. Krauss, and P. Romanelli, Phys. Lett. B 
\textbf{248}, 146 (1990).

\bibitem{Copi} C.J. Copi, A.N. Davis, and L.M. Krauss, astro-ph/0311334.

\bibitem{Olive} K.A. Olive, G. Steigman, and T.P. Walker, Phys. Rep. \textbf{%
333}, 389 (2000).

\bibitem{Kirkman} D. Kirkman, D. Tytler, N. Suzuki, J.M. O'Meara, and D.
Lubin, Ap.J. Suppl. \textbf{149}, 1 (2003).

\bibitem{Kawano} L. Kawano (1992), Fermilab-pub-92/04-A.

\bibitem{webb} V.F. Dmitriev, V.V. Flambaum and J.K. Webb, Phys. Rev. D 
\textbf{69}, 063506 (2004)

\bibitem{IT} Y.I. Izotov and T.X. Thuan, \apj {\bf 500}, 188 (1998).

\bibitem{neutron} G.J. Mathews, T. Kajino, and T. Shima, Phys. Rev. D, submitted, astro-ph/0408523.

\bibitem{Zahn} O. Zahn and M. Zaldarriaga, \prd {\bf 67}, 063002 (2003).

\bibitem{wein} S. Weinberg, Phys. Rev. Lett. \textbf{42}, 850 (1979).

\bibitem{FOT} J. Fry, K. Olive and M.S. Turner, Phys. Rev. D \textbf{22},
2953 (1980).

\bibitem{gib} T. Damour, G. Gibbons and G. Gundlach, Phys. Rev. Lett. 
\textbf{64}, 123 (1990).
\end{thebibliography}
\end{document}